\pdfoutput=1 

\documentclass[
    ,final            
  ]
  {aipproc}

\layoutstyle{6x9}

\begin{document}

\title{Towards a model-independent partial wave analysis for pseudoscalar meson photoproduction}



\author{Lothar Tiator}{
  address={Institut f\"ur Kernphysik, Johannes Gutenberg Universit\"at, D-55099 Mainz, Germany}
}

%

\begin{abstract}
Amplitude and partial wave analyses for pion, eta or kaon photoproduction are
discussed in the context of `complete experiments'. It is shown that the
model-independent helicity amplitudes obtained from at least 8 polarization
observables including beam, target and recoil polarization can not be used to
determine underlying resonance parameters. However, a truncated partial wave
analysis, which theoretically requires only 5 observables will be possible with
minimal model input.
\end{abstract}

\maketitle


\section{Introduction}

Around the year 1970 people started to think about how to determine the four
complex helicity amplitudes for pseudoscalar meson photoproduction from a
complete set of experiments. In 1975 Barker, Donnachie and Storrow published
their classical paper on `Complete Experiments'. After reconsiderations and
careful studies of discrete ambiguities~\cite{FTS92,Keaton:1996pe,chiang}, in
the 90s it became clear that such a model independent amplitude analysis would
require at least 8 polarization observables which have to be carefully chosen.
There are hundreds of possible combinations, but all of them would require a
polarized beam and target and in addition also recoil polarization
measurements. Technically this was not possible until very recently, when
transverse polarized targets came into operation at Mainz, Bonn and JLab and
furthermore recoil polarization measurements by nucleon rescattering has been
shown to be doable.

\section{Complete experiments}

A complete experiment is a set of measurements which is sufficient to predict
all other possible experiments, provided that the measurements are free of
uncertainties. Therefore it is first of all an academical problem, which can be
solved by mathematical algorithms. In practise, however, it will not work in
the same way and either a very high statistical precision would be required,
which is very unlikely, or further measurements of other polarization
observables are necessary. Both problems, first the mathematical problem but
also the problem for a physical experiment can be studied with the help of
state-of-the-art models like MAID or partial wave analyses (PWA) like SAID.
With high precision calculations the complete sets of observables can be
checked and with pseudo-data, generated from models and PWA, real experiments
can be simulated under realistic conditions.

\subsection{Amplitude analysis}

Pseudoscalar meson photoproduction (as $\gamma,\pi$) has 8 spin degrees of
freedom and due to parity conservation it can be described by 4 complex
amplitudes of 2 kinematical variables. Possible sets of amplitudes are:
Invariant amplitudes $A_i$, CGLN amplitudes $F_i$, helicity amplitudes $H_i$ or
transversity amplitudes $b_i$. All of them are linearly related to each other
and further combinations are possible. Most often in the literature the
helicity basis was chosen and the 16 possible polarization observables can be
expressed in bilinear products
\begin{equation}\label{observables}
O_i(W,\theta) =
\frac{q}{k}\,\sum_{k,\ell=1}^4\,\alpha_{k,\ell}\,\,H_k(W,\theta)\,
H_l^*(W,\theta)\,,
\end{equation}
where $O_1$ is the unpolarized differential cross section $\sigma_0$ and all
other observables are products of asymmetries with $\sigma_0$, for details see
Table~\ref{tab:obs}.

From a complete set of 8 measurements $\{O_i(W,\theta)\}$ one can only
determine the moduli of the 4 amplitudes and 3 relative phases. However, there
is always an unknown overall phase, e.g. $\phi_1(W,\theta)\}$, which can not be
determined by additional measurements. Even with the help of unitarity in form
of Watson's  theorem this angle-dependent phase cannot be provided. This has
very strong consequences, namely a partial wave decomposition would lead to
wrong partial waves, which would be useless for nucleon resonance analysis.

\begin{table}
\begin{tabular}{|cc|ccc|c|c|}
\hline
type & $\quad O_i\quad$ & $\quad\alpha_i\quad$ & $\quad\beta_i\quad$ & $\quad\gamma_i\quad$ &  helicity representation    &  Fasano et al.~\cite{FTS92} \\
\hline
         &$\sigma_0$      &      0     &     +1    &     -2     & $\frac{1}{2}(|H_1|^2 + |H_2|^2 + |H_3|^2 + |H_4|^2)$  & $+\sigma_0$           \\
$\mathcal S$ &$\hat{\Sigma}$  &      2     &     -1    &     -2     & Re$(H_1 H_4^* - H_2 H_3^*)$                           & $+\hat{\Sigma}$       \\
         &$\hat{T}$       &      1     &      0    &     -1     & Im$(H_1 H_2^* + H_3 H_4^*)$                           & $+\hat{T}$            \\
         &$\hat{P}$       &      1     &      0    &     -1     & $-$Im$(H_1 H_3^* + H_2 H_4^*)$                        & $+\hat{P}$            \\
\hline
         &$\hat{G}$       &      2     &     -1    &     -1     & $-$Im$(H_1 H_4^* + H_2 H_3^*)$                        & $+\hat{G}$            \\
$\mathcal {BT}$ &$\hat{H}$      &      1     &      0    &     -1     & $-$Im$(H_1 H_3^* - H_2 H_4^*)$                        & $-\hat{H}$            \\
         &$\hat{E}$       &      0     &     +1    &     -1     & $\frac{1}{2}(-|H_1|^2 + |H_2|^2 - |H_3|^2 + |H_4|^2)$ & $-\hat{E}$            \\
         &$\hat{F}$       &      1     &      0    &     -1     & Re$(H_1 H_2^* + H_3 H_4^*)$                           & $+\hat{F}$            \\
\hline
         &$\hat{O_{x'}}$  &      1     &     +1    &     -1     & $-$Im$(H_1 H_2^* - H_3 H_4^*)$                        & $-\hat{O_{x'}}$       \\
$\mathcal {BR}$ &$\hat{O_{z'}}$ &      2     &      0    &     -1     & Im$(H_1 H_4^* - H_2 H_3^*)$                           & $-\hat{O_{z'}}$       \\
         &$\hat{C_{x'}}$  &      1     &     +1    &     -1     & $-$Re$(H_1 H_3^* + H_2 H_4^*)$                        & $-\hat{C_{x'}}$       \\
         &$\hat{C_{z'}}$  &      0     &     +2    &     -1     & $\frac{1}{2}(-|H_1|^2 - |H_2|^2 + |H_3|^2 + |H_4|^2)$ & $-\hat{C_{z'}}$       \\
\hline
         &$\hat{T_{x'}}$  &      2     &      0    &     -2     & Re$(H_1 H_4^* + H_2 H_3^*)$                           & $+\hat{T_{x'}}$       \\
$\mathcal {TR}$ &$\hat{T_{z'}}$ &      1     &     +1    &     -2     & Re$(H_1 H_2^* - H_3 H_4^*)$                           & $+\hat{T_{z'}}$       \\
         &$\hat{L_{x'}}$  &      1     &     +1    &     -2     & $-$Re$(H_1 H_3^* - H_2 H_4^*)$                        & $-\hat{L_{x'}}$       \\
         &$\hat{L_{z'}}$  &      0     &     +2    &     -2     & $\frac{1}{2}(|H_1|^2 - |H_2|^2 - |H_3|^2 + |H_4|^2)$  & $+\hat{L_{z'}}$       \\
\hline
\end{tabular}
\caption{\label{tab:obs}Spin observables expressed by helicity amplitudes in
the notation of Walker~\cite{Walker}. The sign definition is taken from Barker,
Donnachie and Storrow~\cite{Barker75} by replacing $N\rightarrow H_2,\;
S_1\rightarrow H_1,\; S_2\rightarrow H_4,\; D\rightarrow H_3$. This sign
definition is also used by SAID and MAID. The last column compares with Fasano,
Tabakin and Saghai (FTS)~\cite{FTS92}, which is the second most often used sign
definition in the literature. It has been adopted e.g. by recent work of
Anisovich et al.~\cite{Anisovich:2009zy} and Dey et al.~\cite{Dey:2010fb},
while in a paper by Sandorfi et al.~\cite{Sandorfi:2010uv}, a definition very
close to FTS is used, however, with $\hat{E}$ changed into $-\hat{E}$. For the
polarized observables the notation $\hat{\Sigma}:=\sigma_0\,\Sigma$ etc. is
used and a factor $q/k$ is dropped in all observables. Furthermore, the
parameters $\alpha_i,\beta_i,\gamma_i$ for the $cos\theta$ expansions in
Eq.~(\ref{eq:expans0},\ref{eq:expans1}) are given.}
\end{table}

\subsection{Partial wave analysis}

As the main goal in the data analysis of photoproduction is the search for
nucleon resonances and their properties, one can directly perform a partial
wave analysis from the observables without going through the underlying
amplitudes. Such an analysis would be a truncated partial wave analysis with a
minimal model dependence (i) from the truncation of the series at a maximal
angular momentum $\ell_{max}$ and (ii) from an overall unknown phase as in the
case of the amplitude analysis in the previous paragraph. However, in the PWA
the overall phase would only be a function of energy and with additional
theoretical help it can be constrained without strong model assumptions. Such a
concept was already discussed and applied for $\gamma,\pi$ in the 80s by
Grushin~\cite{grushin} for a PWA in the region of the $\Delta(1232)$ resonance.
In a 2-step process first $\gamma,\pi^+$ was analyzed in a truncation to $S$
and $P$ waves, where all higher partial waves were taken as Born terms,
dominantly from the pion-pole contribution, which contributes only to charged
pion production. This had the additional advantage that the overall phase was
determined by the real Born amplitudes for $\ell>1$. In order to separate the
isospin 1/2 and 3/2 a further analysis has to be done for $\gamma,\pi^0$. For
neutral pion production, the higher partial waves are much smaller and can be
neglected, however, in this case the overall phase remains undetermined. This
problem could be solved by applying the Watson theorem just for the $P_{33}$
partial wave in the following way~\cite{grushin,Workman:2010xc}
\begin{equation}
M_{1+}^{\pi^0p} =
\alpha\,e^{i\,\delta_{33}}\,+\,\frac{1}{\sqrt{2}}\,M_{1+}^{\pi^+n}\,.
\label{eq:phase}
\end{equation}

Formally, the truncated partial wave analysis can be performed in the following
way. All observables can be expanded either in a Legendre series or in a
$cos\theta$ series, the first one has no real advantage, as we will see there
is no physical angular momentum involved in the order of the power series
\begin{eqnarray}
O_i(W,\theta) &=& \frac{q}{k}\,\,
sin^{\alpha_i}\theta\,\sum_{k=0}^{2\ell_{max}+\beta_i}
a_k^i(W)\,\,cos^k\theta\,, \label{eq:expans0}\\
a_k^i(W) &=& \sum_{\ell,\ell'=0}^{\ell_{max}}\; \sum_{k,k'=1}^4
\alpha_{\ell,\ell'}^{k,k'}\,\mathcal M_{\ell,k}(W)\, \mathcal
M_{\ell',k'}^*(W)\,, \label{eq:expans0b}
\end{eqnarray} where $k,k'$ denote the 4
possible electric and magnetic multipoles for each $\pi N$ angular momentum
$\ell$, namely $\mathcal
M_{\ell,k}=\{E_{\ell+},E_{\ell-},M_{\ell+},M_{\ell-}\}$. For an $S,P$ wave
truncation $(\ell_{max}=1)$ there are 4 complex multipoles
$E_{0+},E_{1+},M_{1+},M_{1-}$ leading to 7 free real parameters and an
arbitrary phase, which can be put to zero for the beginning. With the expansion
parameters $\alpha_i,\beta_i$ from Table~\ref{tab:obs} one can already get 8
observable coefficients from the first group of 4 $\mathcal S$-type
observables. This already exceeds the number of free parameters, however, in
order to resolve discrete ambiguities, one more observable or at least one more
coefficient needs to be measured. This can be taken from any type of double
polarization $\mathcal {BT},\mathcal {BR},\mathcal {TR}$. As has been shown by
Omelaenko~\cite{omel} the same is true for any PWA with truncation at
$\ell_{max}$. For the determination of the $8\ell_{max}-1$ free parameters one
has the possibility to measure
$(8\ell_{max},\,8\ell_{max},\,8\ell_{max}+4,\,8\ell_{max}+4)$ coefficients for
types $(\mathcal S,\mathcal {BT},\mathcal {BR},\mathcal {TR})$, respectively.

However, this is only the mathematical issue of the problem. For any finite
precision of data, the solution of the problem becomes more involved and
instead of exact solutions of the sets of quadratic equations one has to search
for a minimum similar to a chisquare fit. Then the global minimum, which may
already be difficult to find does not necessarily give the correct solution.
Therefore also local minima have to be considered and further techniques are
needed to arrive at the correct solution. On the experimental side one has two
possibilities to improve the situation, either to get the observables at higher
precision or to measure further observables than mathematically required.

In Ref.~\cite{Sandorfi:2010uv} in an analysis of kaon photoproduction with real
world data such a problem with multiple local minima has been faced and work is
in progress. Similar problems will probably occur also in eta photoproduction,
however, in pion photoproduction it will be different. There, the more
difficult task to measure and analyze 2 charged channels, $p(\gamma,\pi^0)p$
and $p(\gamma,\pi^+)n$ also provides a chance to solve the problem with only a
minimal model input. The big advantage in this reaction is the existence of the
pion-pole term, which contributes only to charged pion channels and which gives
large contributions for higher partial waves. Since the pion-nucleon coupling
constant is well known, these higher partial waves can be very precisely
calculated and furthermore their phases are known, namely zero for real Born
terms. This fixes the overall phase for $\pi^+n$ partial waves. In a subsequent
analysis of $\pi^0$ photoproduction these higher Born terms do not contribute,
but as mentioned before, the overall phase can then be connected to the
$\pi^+n$ phase, Eq.~(\ref{eq:phase}). Of course this connection needs to be
done only for one phase, which naturally will be taken as the $P_{33}$ phase,
which is very well known and which is elastic even up to $W\approx 1550$~MeV,
whereas other phases as $P_{11}$ become already inelastic for $W\approx
1300$~MeV, close to the $\pi\pi$ threshold. This method was first applied in
the analysis of Grushin~\cite{grushin}.

In the presence of the $t$-channel pole contribution, the expansion of
Eq.~(\ref{eq:expans0}) must be modified by
\begin{eqnarray}
O_i(W,\theta) &=& \frac{q}{k}\,\,
sin^{\alpha_i}\theta\,\sum_{k=\gamma_i}^{2\ell_{max}+\beta_i}
a_k^i(W)\,(1-\frac{q_{\pi^+}}{\omega_{\pi^+}}cos\theta)^k\,, \label{eq:expans1}
\end{eqnarray} which is an expansion around the pion pole.
From the values of $\gamma_i$ in Table~\ref{tab:obs} one can see that most
observables start with a single-pole structure $\sim 1/\kappa$ with
$\kappa=1-q/\omega\,cos\theta$, whereas the unpolarized cross section
$\sigma_0$ and the beam asymmetry $\hat{\Sigma}$ as well as all observables of
$\mathcal {BR}$ type include an additional double-pole term $\sim 1/\kappa^2$,
which is, however, completely fixed by the Born terms, hence by the $\pi N$
coupling constant. Therefore, only the coefficients $a_{-1}^i(W)$ have to be
obtained from the angular distributions of the observables.

\section{Partial wave analysis with pseudo-data}

In a first numerical attempt towards a model-independent partial wave analysis,
a procedure similar to the second method described above has been
applied~\cite{Workman:2011hi}, and pseudo-data, generated for $\gamma,\pi^0$
and $\gamma,\pi^+$ have been analyzed.

Events were generated over an energy range from $E_{lab}=200-1200$~MeV and a
full angular range of $\theta=0-180^\circ$ for beam energy bins of $\Delta
E_{\gamma}=10$~MeV and angular bins of $\Delta\theta=10^\circ$, based on the
MAID2007 model predictions~\cite{Drechsel:2007if}. For each observable,
typically $5\cdot10^6$ events have been generated over the full energy range.
\begin{figure}
\includegraphics[width=0.36\textwidth, keepaspectratio, angle=90]{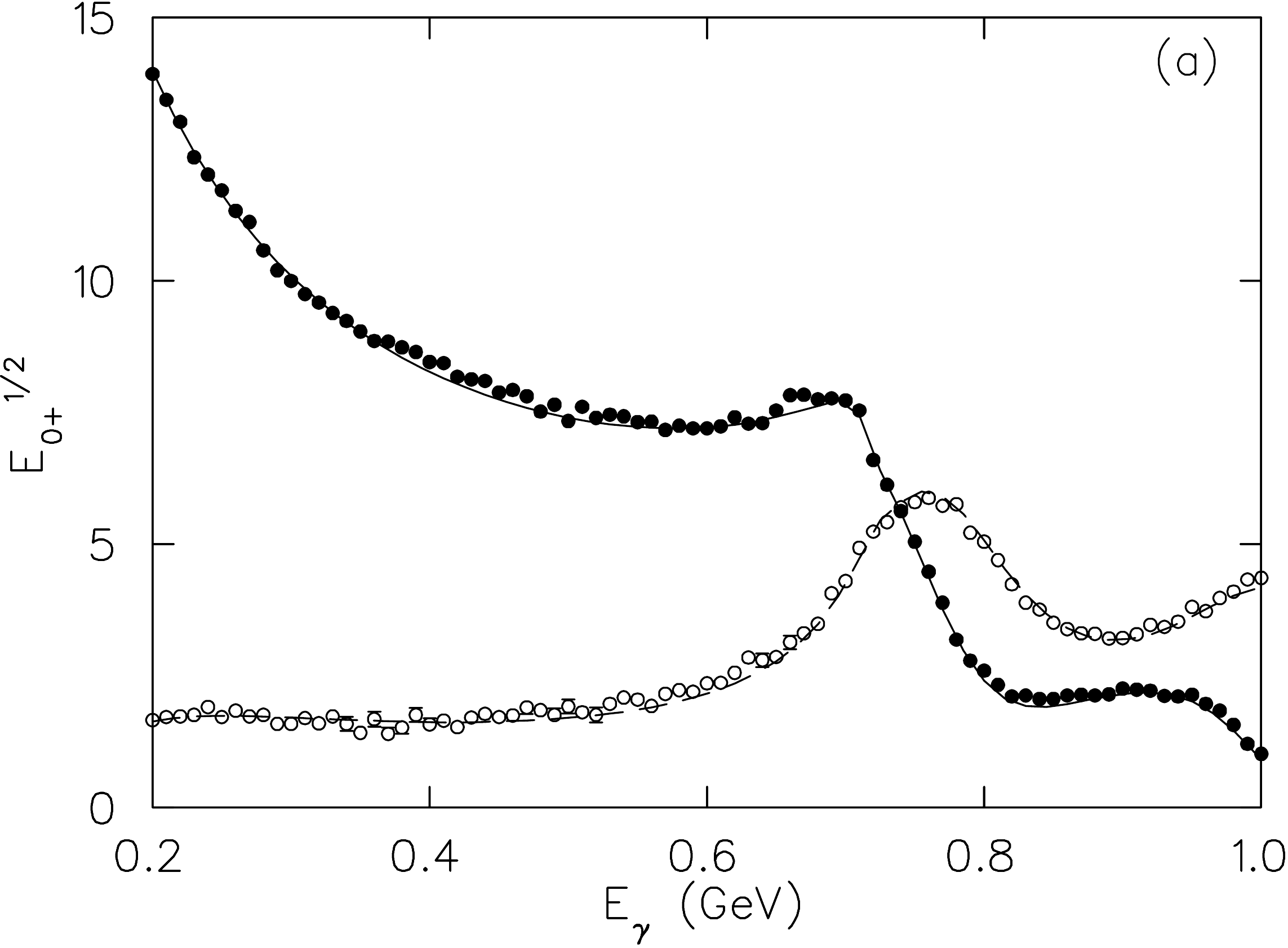}
\includegraphics[width=0.36\textwidth, keepaspectratio, angle=90]{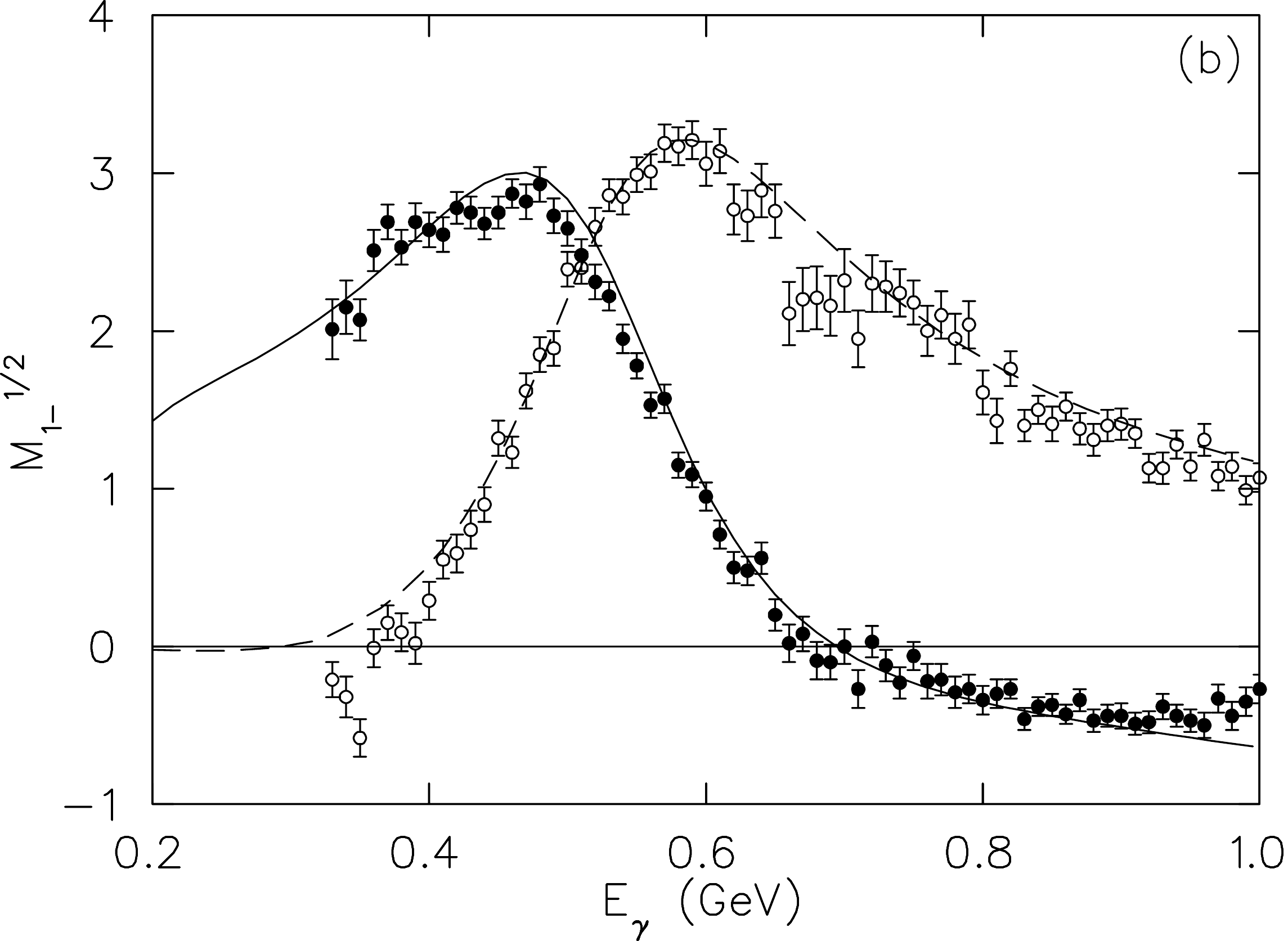}
\caption{\label{fig:p11s11} Real and imaginary parts of (a) the $S_{11}(pE)$
partial wave amplitude $E_{0+}^{1/2}$ and (b) the $P_{11}(pM)$ partial wave
amplitude $M_{1-}^{1/2}$. The solid (dashed) line shows the real (imaginary)
part of the MAID2007 solution, used for the pseudo-data generation.  Solid
(open) circles display real (imaginary) single-energy fits including the final
phase searching (SE6p) to the following 6 observables without any recoil
polarization measurement: $d\sigma/d\Omega$, two single-spin observables
$\Sigma$ and $T$ and three beam-target double polarization observables $E$, $F$
and $G$.}
\end{figure}
For each energy bin a single-energy (SE) analysis has been performed using the
SAID PWA tools~\cite{Arndt:1989ww} in a 3-step process:
\begin{enumerate}
  \item An energy-dependent analysis (ED) was obtained from starting
  values given by the current SAID SP09 ED solution.

  \item By fixing the phases of the multipoles to the ED solution, only the
  absolute values of the $S,P,D$ and $F$ multipoles ($\ell_{max}$ depends on the
  energy) were searched.

  \item In a final step also the phases of the multipoles were searched and a very
  satisfactory result was found in comparison with the underlying MAID2007 solution.
\end{enumerate}

A series of fits have been performed~\cite{Workman:2011hi} using 4, 6 and 8
observables. Here the example using 6 observables $(\sigma_0,\Sigma,T,E,F,G)$
is demonstrated, where no recoil polarization has been used. As explained
before, such an experiment would be incomplete in the sense of an `amplitude
analysis', but complete for a truncated partial wave analysis. In
Fig.~\ref{fig:p11s11} two multipoles $E_{0+}^{1/2}$  and $M_{1-}^{1/2}$ for the
$S_{11}$ and $P_{11}$ channels are shown and the SE6p fits of the final step in
the analysis are compared to the MAID2007 solution. The fitted SE solutions are
very close to the MAID ED solution with very small uncertainties for the
$S_{11}$ partial wave. For the $P_{11}$ partial wave we obtain a larger
statistical spread of the SE solutions. This is typical for the $M_{1-}^{1/2}$
multipole, which is generally much more difficult to obtain with good
accuracy~\cite{Arndt:2002xv,Drechsel:2007if}, because of the weaker sensitivity
of the observables to this magnetic multipole. But also this multipole can be
considerably improved in an analysis with 8 observabes~\cite{Workman:2011hi}.

\section{Summary and conclusions}

It is shown that for an analysis of $N^*$ resonances, the amplitude analysis of
a complete experiment is not very useful, because of an unknown energy and
angle dependent phase that can not be determined by experiment and can not be
provided by theory without a strong model dependence. However, the same
measurements will be very useful for a truncated partial wave analysis with
minimal model dependence due to truncations and extrapolations of Watson's
theorem in the inelastic energy region. A further big advantage of such a PWA
is a different counting of the necessary polarization observables, resulting in
very different sets of observables. While it is certainly helpful to have
polarization observables from 3 or 4 different types of Table~\ref{tab:obs},
for a mathematical solution of the bilinear equations one can find minimal sets
of only 5 observables from only 2 types, where either a polarized target or
recoil polarization measurements can be completely avoided.


\begin{theacknowledgments}
I would like to thank R. Workman, M. Paris, J. Gegelia, S. Kamalov, M. Ostrick
and S. Schumann for helpful discussions and contributions to this ongoing work.
This work is supported by the Deutsche Forschungsgemeinschaft (SFB443).
\end{theacknowledgments}

%
\end{document}